# Operations on polytopes: application to tolerance analysis

*TEISSANDIER Denis\*, DELOS Vincent\*\*, COUETARD Yves\**
*\*LMP – Laboratoire de Mécanique Physique – UPRES A 5469 CNRS*
*Université Bordeaux 1*
*351, Cours de la Libération*
*33405 Talence Cedex, France*
*Phone : 05 56 84 62 20, Fax : 05 56 84 69 64*
*Email : d.teissandier@lmp.u-bordeaux.fr, couetard@lmp.u-bordeaux.fr*

*\*\*BRGM, Avenue de Concyr, Orléans-La-Source - BP 6009*
*45060 Orléans Cedex 2 - France*
*Phone : 33-(0)3-38-64-34-34 - fax : 33-(0)3-38-64-35-18*

**Abstract :** This article presents numerical methods in order to solve problems of tolerance analysis. A geometric specification, a contact specification and a functional requirement can be respectively characterized by a finite set of geometric constraints, a finite set of contact constraints and a finite set of functional constraints. Mathematically each constraint formalises a n-face (hyperplan of dimension n) of a n-polytope ($1 \leq n \leq 6$). Thus the relative position between two any surfaces of a mechanism can be calculated with two operations on polytopes : the Minkowski sum and the Intersection. The result is a new polytope: the calculated polytope. The inclusion of the calculated polytope inside the functional polytope indicates if the functional requirement is satisfied or not satisfied. Examples illustrate these numerical methods.

1.  INTRODUCTION

The variational classes by Requicha were introduced at the beginning of the 80's and propose a model of tolerances of form, orientation, position and dimension [Requicha, 1983]. A generalization of specifications by volume envelopes is based on the works of Requicha [Srinivasan and al., 1989]. Fleming presents a model for geometric tolerances and constraints from contacts [Fleming, 1988]. Among the dimension-chains models based on Small Displacements Torsor concept [Clément and al., 1988], we note the Clearance Deviation Space by Giordano [Giordano and al., 1993]. The Clearance Deviation Space purposes an assembly method according to maximum material condition limit.

This article presents numerical methods in order to solve problems of tolerance analysis. These methods deal with Tolerance Zones constructed by offsetting as in [Requicha, 1983]. We use the Small Displacements Torsor [Bourdet and al., 1995]. Geometric specification and contact specification are characterised by the same model as in [Giordano and al., 1993].

2. EXPRESSION OF CONSTRAINTS

A Tolerancing Tool manipulates three sources of essential information:
a. the geometric specifications (between associated surfaces of the same part).
b. the contact specifications (between associated surfaces of two distinct parts),
c. the functional requirements of an assembly (between any associated surfaces).

### 2.1. Geometric constraints

An associated surface is a surface of perfect form (i.e. an ideal surface: surface described with a finite number of geometric features). A nominal surface is an ideal surface by definition.
A geometric specification is formalised by geometric constraints of position between a nominal surface $S_0$ and an associated surface $S_1$: see figure 1.
The tolerance zone (ZT) limit an area of space around $S_0$ within which $S_1$ must be situated: they are constructed by two (positive and negative) offsettings on $S_0$.

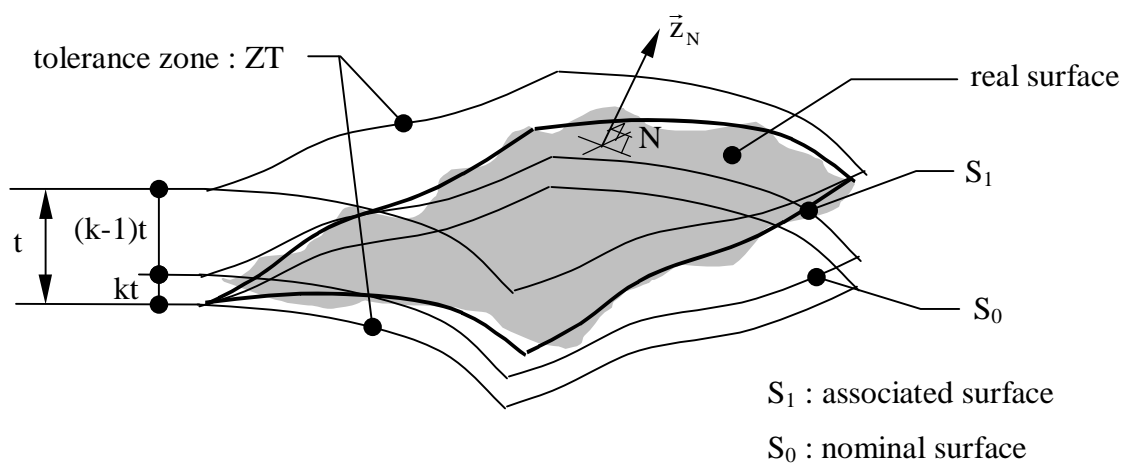

*Figure 1: Geometric specification - geometric constraints.*

We define geometric constraints of position between $S_1$ and $S_0$ as follows [Teissandier et al., 1997]:

$$S_1 \subset ZT \Leftrightarrow \forall N \in S_0 \quad \text{with } 0 \leq k \leq 1 \qquad (k-1).t \leq \vec{\varepsilon}_{N,1/0}.\vec{z}_N \leq k.t \qquad (1)$$

A unit vector $\vec{z}_N$ is constructed such as $\vec{z}_N$ is parallel to the local normal at any point N of $S_0$ (see figure 1). Vector $\vec{z}_N$ is oriented in such a way that the positive direction corresponds to the side exterior of the material.

With:
$$\begin{cases} \vec{\varepsilon}_{N,1/0} : \text{translation vector between } S_1 \text{ and } S_0 \text{ expressed at point N.} \\ \vec{\rho}_{1/0} : \text{rotation vector between } S_1 \text{ and } S_0. \\ \forall M \in E^3 \ (E^3 : \text{Euclidean Space}) \quad \vec{\varepsilon}_{N,1/0} = \vec{\varepsilon}_{M,1/0} + \vec{\rho}_{1/0} \times \overrightarrow{MN} \end{cases}$$

We use the property of linearization of displacements into small displacements [Bourdet et al., 1995], [Clément et all., 1988].
We get according to (1):

$$S_1 \subset ZT \Leftrightarrow \forall N \in S_0, \forall M \in E^3 \text{ with } 0 \leq k \leq 1 \quad -kt \leq \left(\vec{\varepsilon}_{M,1/0} + \vec{\rho}_{1/0} \times \overrightarrow{MN}\right)\vec{z}_N \leq (1-k)t \quad (2)$$

In a base $(\vec{x}, \vec{y}, \vec{z})$, the normal vector $\vec{z}_N$ of $S_0$ at point N can be written as follows:

$$\vec{z}_N \begin{pmatrix} a_N \\ b_N \\ c_N \end{pmatrix} \text{ and } \overrightarrow{MN} \begin{pmatrix} d_{MNx} \\ d_{MNy} \\ d_{MNz} \end{pmatrix}.$$

$(2) \Rightarrow \quad S_1 \subset ZT \Leftrightarrow \forall N \in S_0, \quad \forall M \in E^3 \text{ with } 0 \leq k \leq 1 :$

$$(k-1)t \leq \left(\varepsilon_{1/0x,M} + \rho_{1/0y} \cdot d_{MNz} - \rho_{1/0z} \cdot d_{MNy}\right)a_N + \left(\varepsilon_{1/0y,M} - \rho_{1/0x} \cdot d_{MNz} + \rho_{1/0z} \cdot d_{MNx}\right)b_N \quad (3)$$
$$+ \left(\varepsilon_{1/0z,M} + \rho_{1/0x} \cdot d_{MNy} - \rho_{1/0y} \cdot d_{MNx}\right)c_N \leq k.t$$

We obtain an infinity of equations (3). The unknowns are the six components written at point M:

$\rho_{1/0x}, \quad \rho_{1/0y}, \quad \rho_{1/0z}, \quad \varepsilon_{M,1/0y}, \quad \varepsilon_{M,1/0z}, \quad \varepsilon_{M,1/0z}.$

Any surface can be discretised into n points $N_i$. So, it is possible to express a set of n equations (3). The n equations (3) characterize the n geometric constraints induced by the tolerance zone associated with $S_0$.

The vertices of this polytope correspond to the maximum and the minimum values of

$\rho_{1/0x}, \quad \rho_{1/0y}, \quad \rho_{1/0z}, \quad \varepsilon_{M,1/0y}, \quad \varepsilon_{M,1/0z}, \quad \varepsilon_{M,1/0z}.$

This method can be applied on any ideal surface. We consider five types of surfaces: plane, cylindrical, conic and toric surfaces.

### 2.2. Contact constraints

Amongst the five main types of surfaces considered in the previous paragraph : plane, cylindrical, conic and toric surfaces, figure 2 summarizes the possible cases of joint. Since complex surfaces are not used in a joint between two parts (with the exception of a few particular cases i.e. gearing ) they will not be considered. Each case in the above table can be sub-classified into several other cases according to the relative position of nominal surfaces. Any joint is characterized by the types of two considered surfaces with a set of mating conditions. A mating condition is a set of constraints between geometric features of two ideal surfaces [Teissandier, 1995]. An exhaustive list of the cases has been compiled by [Clément et al., 1997]: in order to define the relative position between two any surfaces 13 constraints has been identified.

|        | plane | cylinder | cone | sphere | tore |
|--------|-------|----------|------|--------|------|
| plane  | ×     | ×        | ×    | ×      | ×    |
| cylinder |     | ×        | ×    | ×      | ×    |
| cone   |       |          | ×    | ×      | ×    |
| sphere |       |          |      | ×      | ×    |
| tore   |       |          |      |        | ×    |

*Figure 2: Main cases of joints.*

Following the same method as in the previous paragraph, a contact specification can be formalised by a set of n contact constraints of position between two associated surfaces of two distinct parts. Let us consider a joint made up of two planes $S_1$ and $S_2$. $S_1$ and $S_2$ are nominally parallel and separated by a distance « d » (see figure 3). Vectors $\vec{z}_1$ and $\vec{z}_2$ are respectively constructed such as $\vec{z}_1$ and $\vec{z}_2$ are normal vectors of $S_1$ and $S_2$ (see figure 3) oriented in such a way that the positive direction corresponds to the side exterior to the material.

Let us define surface S such as : $S = S_1 \cap S_2$ with $d = 0$. The set of mating conditions is:

$$\{\vec{z}_1 \times \vec{z}_2 = \vec{0}, \quad \vec{z}_1 \cdot \vec{z}_2 < 0, \quad S \text{ is a plane surface}\} \quad (4)$$

If the set of mating conditions is satisfied, we can express a constraint of positioning : $0 \leq d \leq D$ (5)

A permanent contact between $S_1$ and $S_2$ is such as: $D = 0 \Rightarrow 0 \leq d \leq 0 \Rightarrow d = 0$.

If S is not a plane (line, point or empty hole), the previous constraint can not be defined: the set of mating condition is not verified.

Let us consider the boundary (C) of S.

At any point M of the Euclidean space $E^3$, it is therefore:

$$\forall N \in S, \forall M \in E^3 \; (4, 5) \Rightarrow \left(\vec{\varepsilon}_{M,1/2} + \vec{\rho}_{1/2} \times \overrightarrow{MN}\right) \vec{z}_N \leq d \Rightarrow 0 \leq \left(\vec{\varepsilon}_{M,1/2} + \vec{\rho}_{1/2} \times \overrightarrow{MN}\right) \vec{z}_N \leq D \quad (6)$$

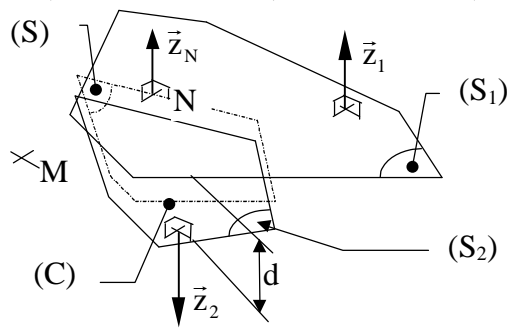

*Figure 3: Contact specification between two nominal parallel planes.*

Following the same method as geometric constraints, we can write:

$$\forall N \in S, \quad \forall M \in E^3 \; : \; (4, 5, 6) \Rightarrow 0 \leq \left(\varepsilon_{1/0x,M} + \rho_{1/0y} \cdot d_{MNz} - \rho_{1/0z} \cdot d_{MNy}\right) a_N +$$
$$\left(\varepsilon_{1/0y,M} - \rho_{1/0x} \cdot d_{MNz} + \rho_{1/0z} \cdot d_{MNx}\right) b_N + \left(\varepsilon_{1/0z,M} + \rho_{1/0x} \cdot d_{MNy} - \rho_{1/0y} \cdot d_{MNx}\right) c_N \leq D \quad (7)$$

If $\vec{z}_N = \vec{z}$, this corresponds to the particular case where: $a_N = b_N = 0$ and $c_N = 1$.
Thus (7) can be written as follows:

$$\forall N \in S, \quad \forall M \in E^3 : (4, 5, 6) \Rightarrow 0 \leq \varepsilon_{1/0z,M} + \rho_{1/0x} \cdot d_{MNy} - \rho_{1/0y} \cdot d_{MNx} \leq D \qquad (8)$$

Contact constraints (8) traduce the three degrees of freedom of the studied joint: 1 displacement in rotation and 2 displacements in translation at point M. As in the previous paragraph, we can obtain a finite set of n equations (7).

### 2.3. Functional constraints

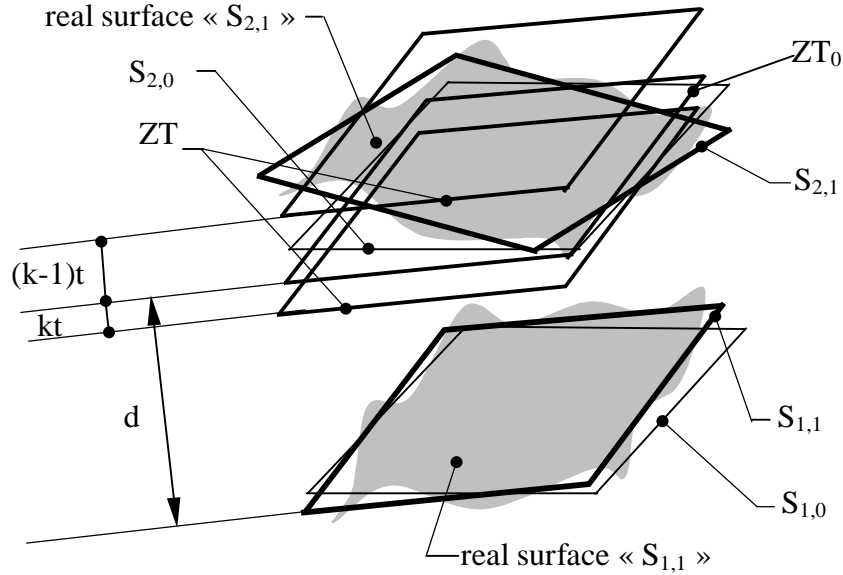

*Figure 4: Functional requirement.*

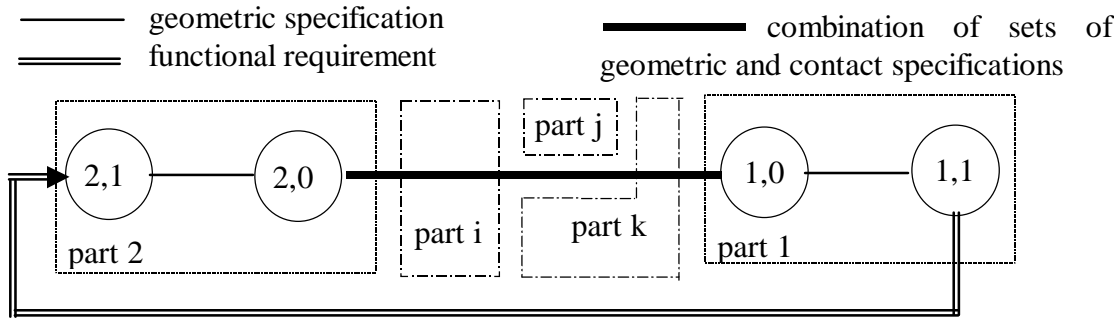

*Figure 5: Surface graph according to figure 4.*

The figure 4 illustrates an example of two surfaces $S_{2,1}$ and $S_{1,1}$ of any mechanism. $S_{2,1}$ is an associated surface of nominal surface $S_{2,0}$ and $S_{1,1}$ is an associated surface of nominal surface $S_{1,0}$: see figure 4. The relative position between $S_{2,1}$ and $S_{1,1}$ is the result of a combination of sets of contact specifications and geometric specifications on different parts: see figures 4 and 5. In the figure 4 $S_{2,0}$ and $S_{1,0}$ are two nominal parallel planes. A functional requirement between $S_{2,1}$ with regards to $S_{1,1}$ is defined by a tolerance zone ZT which limits an area of space around $ZT_0$ within which $S_{2,1}$ must be situated: see figure 4. ZT is composed of two parallel planes separated by the dimension t of ZT. These two surfaces are parallel to $ZT_0$, a parallel plane to $S_{1,1}$. The relative position between $S_{1,1}$ and $ZT_0$ is specified by the dimension d. A functional requirement is formalised by a finite set of n functional constraints between any associated surface with regards to any surface of the same mechanism. The method is the same as geometric specification.

$S_{2,1} \subset ZT \Leftrightarrow \forall N \in S_{2,0}, \quad \forall M \in E^3 \text{ avec } 0 \leq k \leq 1 : (k-1)t \leq \vec{\varepsilon}_{2,1/1,1x,M} \cdot \vec{z}_N \leq k.t \Rightarrow$

$(k-1)t \leq (\varepsilon_{2,1/1,1x,M} + \rho_{2,1/1,1y} \cdot d_{MNz} - \rho_{2,1/1,1z} \cdot d_{MNy}) a_N +$  (9)

$(\varepsilon_{2,1/1,1y,M} - \rho_{2,1/1,1x} \cdot d_{MNz} + \rho_{2,1/1,1z} \cdot d_{MNx}) b_N + (\varepsilon_{2,1/1,1z,M} + \rho_{2,1/1,1x} \cdot d_{MNy} - \rho_{2,1/1,1y} \cdot d_{MNx}) c_N \leq k.t$

We obtain a finite set of n equations (8). The surfaces $S_{1,1}$ and $S_{2,1}$ can be (or can not be) specified from the same part. That means that a functional requirement can be reduced to a combination of two geometric specifications.

### 3. OPERATIONS ON POLYTOPES

#### 3.1. Definition of a polytope

Geometric specifications, contact specifications and functional requirements can be respectively characterized by a finite set of geometric constraints (3), a finite set of contact constraints (8) and a finite set of functional constraints (9).

Each constraint of (3), (8) and (9) corresponds to a n-face (hyperplan of dimension n: $0 \leq n \leq 6$) in the real affine space $R^d$. Mathematically, (3) defines a geometric n-polytope (polytope of dimension n). By analogy, (8) defines a contact n-polytope and (9) defines a functional n-polytope.

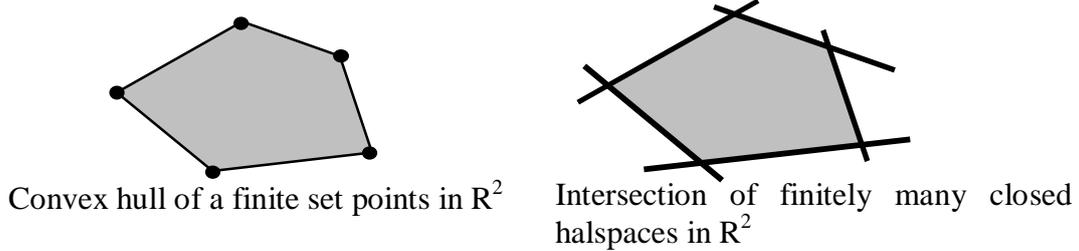

Convex hull of a finite set points in $R^2$     Intersection of finitely many closed halspaces in $R^2$

*Figure 6: Definition of a 2-polytope in $R^2$.*

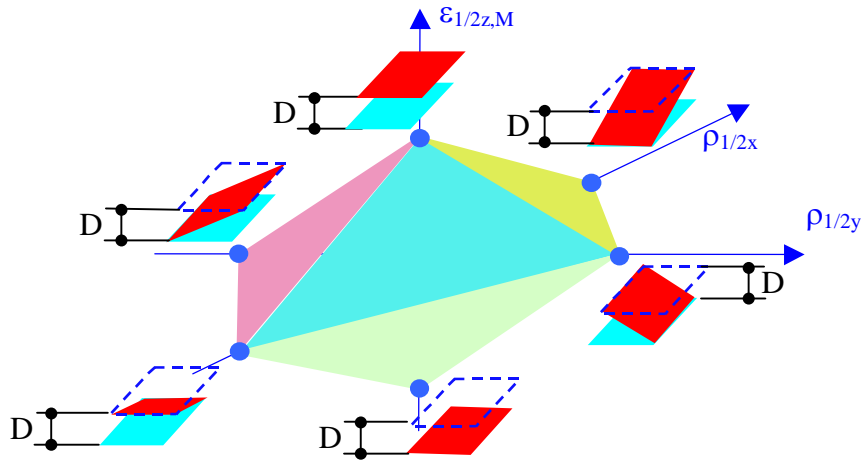

***Figure 7:*** *3-polytope in $R^3$ from contact specification between two rectangular planes.*

A polytope is a point set $P \subseteq R^d$ which can be presented either as [Ziegler, 1995]:
- a bounded intersection of finitely many closed halfspaces in some $R^d$,
- a convex hull of a finite set of points in some $R^d$.

The figure 6 illustrates the two definitions of a polytope. The dimension of a polytope is the dimension of its affine hull. A n-polytope is a polytope of dimension n in some $R^d$. For example, we have defined a finite set of contact constraints (7) specified on two nominal parallel planes. (7) is a finite set of 3-faces in $R^6$ and (8) is a finite set of 3-faces in $R^3$. It is possible to give a graphic representation of a n-polytope if $1 \leq n \leq 3$. A 0-polytope is a point, a 1-polytope is a segment line and a 2-polytope is a polygon. For example, the finite set of contact constraints (7) (i.e. finite set of 3-faces) between two nominal parallel planes such as S is rectangular plane can be illustrated in $R^3$ by figure 7.

### 3.2. Minkowski sums of polytopes

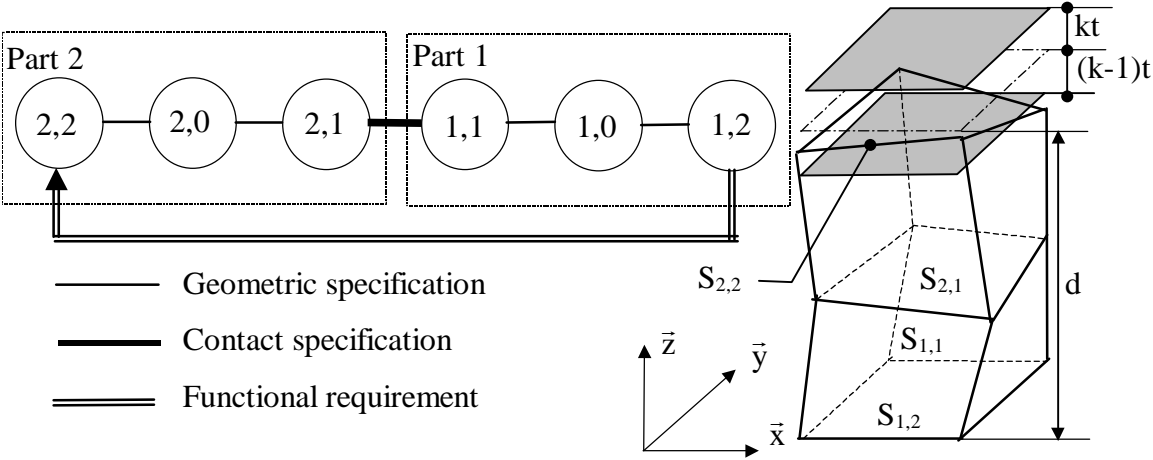

***Figure 8:*** *Association of specifications in series.*

Let us consider the mechanism of figure 8. The contact specification between $S_{2,1}$ and $S_{1,1}$ is a permanent contact. The corresponding contact polytope is a 0-polytope (i.e. a point). We can express the relative position of $S_{2,2}$ with regards to $S_{1,2}$. We have according to the properties of small displacements:

$$\forall N \in S_{2,2} \quad S_{2,2} \subset ZT \Leftrightarrow (k-1).t \leq \vec{\varepsilon}_{N,S2,2/S1,2}.\vec{z}_N \leq k.t \tag{10}$$

$$\left.\begin{array}{l}\vec{\varepsilon}_{M,2,2/1,2} = \vec{\varepsilon}_{M,2,2/2,0} + \vec{\varepsilon}_{M,2,0/2,1} + \vec{\varepsilon}_{M,2,1/1,1} + \vec{\varepsilon}_{M,1,1/1,0} + \vec{\varepsilon}_{M,1,0/1,2} \\ \forall N \in S_{2,2} \text{ with } 0 \leq k \leq 1 : (k-1)t_{2,2} \leq \vec{\varepsilon}_{N,2,2/2,0}.\vec{z} \leq k.t_{2,2} \\ \forall N \in S_{2,1} \text{ with } 0 \leq k \leq 1 : (k-1)t_{2,1} \leq \vec{\varepsilon}_{N,2,1/2,0}.(-\vec{z}) \leq k.t_{2,1} \\ \forall N \in S_{2,1/1,1} \quad \vec{\varepsilon}_{N,1,2/2,2}.\vec{z} = 0 \\ \forall N \in S_{1,2} \text{ with } 0 \leq k \leq 1 : (k-1)t_{1,2} \leq \vec{\varepsilon}_{N1,2/1,0}.(-\vec{z}) \leq k.t_{1,2} \\ \forall N \in S_{1,1} \text{ with } 0 \leq k \leq 1 : (k-1)t_{1,1} \leq \vec{\varepsilon}_{N1,1/1,0}.\vec{z} \leq k.t_{1,1}\end{array}\right\} \tag{11}$$

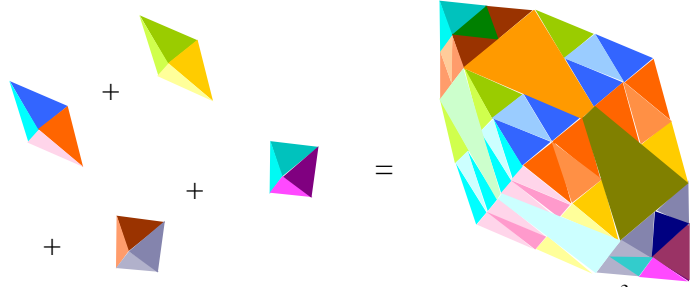

*Figure 9: Minkowski sum of 3-polytopes in $R^3$.*

(11) characterizes the Minkowski sum of five polytopes. These five polytopes are the geometric specifications of $S_{2,2}$, $S_{2,1}$, $S_{1,1}$, $S_{1,2}$ and the contact polytope between $S_{1,2}$ and $S_{2,2}$ (in this case this polytope is a 0-polytope). The Minkowski sum of a two polytopes $P_1$ and $P_2$ is a polytope $P_1 + P_2$ [Gritzmann et al., 1993]:

$$P_1 + P_2 = \{x \in R^d / \exists x_1 \in P_1, \exists x_2 \in P_2 : x = x_1 + x_2\} \tag{12}$$

The association of geometric specifications and contact specifications in series are mathematically formalised by Minkowski sums of d-polytopes [Srinivasan, 1993]. In the example of figure 8, we can illustrate the Minkowski sum of five 3-polytopes in $R^3$: see figure 9. The Minkowski sum is an commutative and associative operation.

### 3.3. Intersection of polytopes

In the example presented in figures 10 and 11, the relative position between $S_{1,0}$ and $S_{2,0}$ must satisfy the two following relations:

$$\forall M \in E^3 \quad \begin{cases} \vec{\varepsilon}_{M,2,0/1,0} = \vec{\varepsilon}_{M,2,0/2,1} + \vec{\varepsilon}_{M,2,1/1,1} + \vec{\varepsilon}_{N,1,1/1,0} \\ \vec{\varepsilon}_{M,2,0/1,0} = \vec{\varepsilon}_{M,2,0/2,2} + \vec{\varepsilon}_{M,2,2/1,2} + \vec{\varepsilon}_{N,1,2/1,0} \end{cases} \tag{13}$$

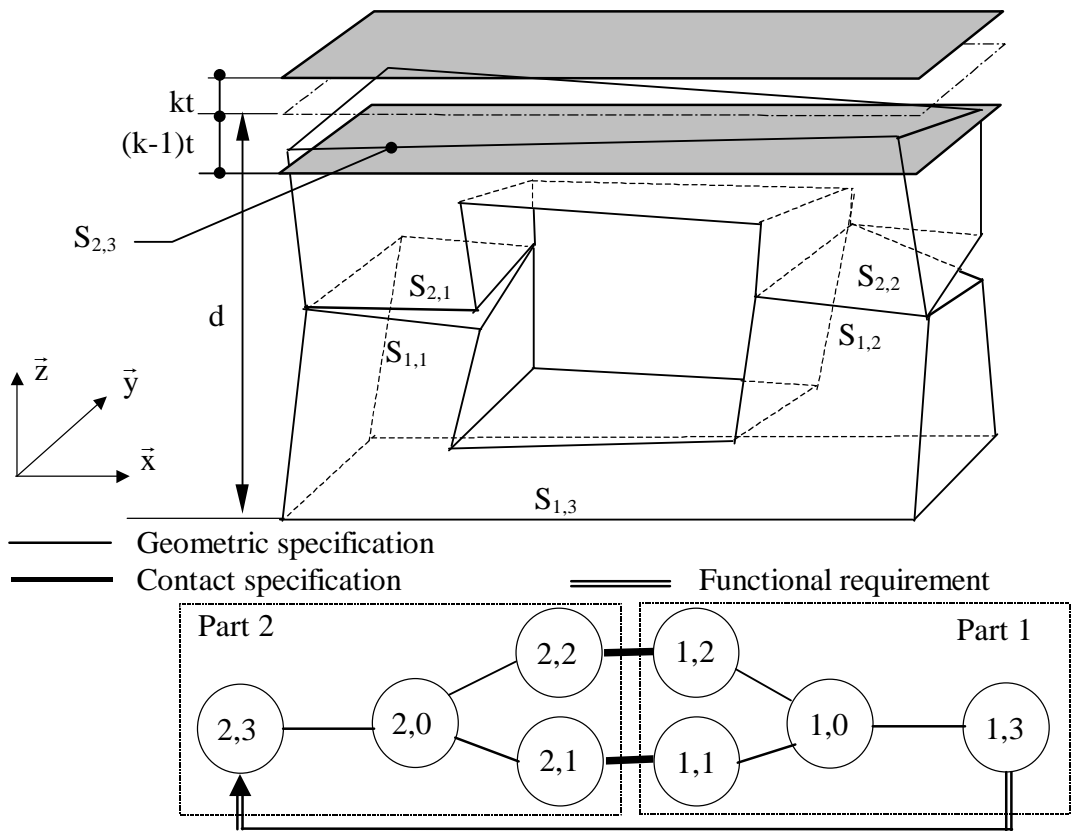

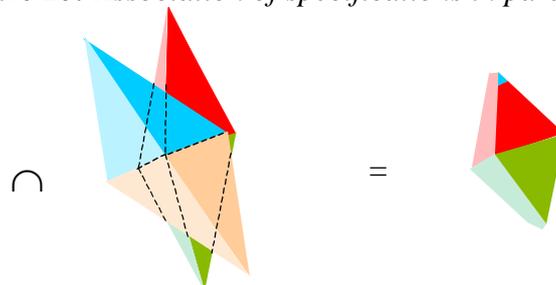

*Figure 10: Association of specifications in parallel.*

*Figure 11: Intersection of 3-polytopes.*

Each relation of (12) can be formalised by a Minkowski sum of three 3-polytopes. The result of (12) is an Intersection of two Minkowski sums of 3-polytopes: see figure 11.

### 4. CONCLUSION: TOLERANCING ANALYSIS WITH POLYTOPES

With Minkowski sums and Intersections of geometric d-polytopes and contact d-polytopes, the relative position between any surfaces of a mechanism can be calculated. The result is a new polytope: the calculated polytope. If the calculated polytope is included inside the functional polytope, the functional requirement is satisfied: see figure 12.

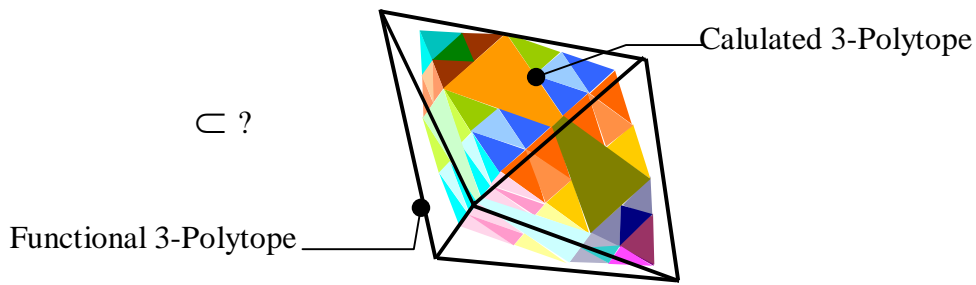

*Figure 12: Verification of inclusion of 3-polytope.*